\documentclass[12pt,a4paper]{article}

\usepackage{essv}
\usepackage{expdlist}

\title{More than words: Advancements and Challenges in Speech Recognition for Singing}
\author{Anna Kruspe}
\affil{Munich University of Applied Sciences}
\email{anna.kruspe@hm.edu}

\begin{document}


\maketitle

\begin{abstract}
  This paper addresses the challenges and advancements in speech recognition for singing, a domain distinctly different from standard speech recognition. Singing encompasses unique challenges, including extensive pitch variations, diverse vocal styles, and background music interference. We explore key areas such as phoneme recognition, language identification in songs, keyword spotting, and full lyrics transcription. I will describe some of my own experiences when performing research on these tasks just as they were starting to gain traction, but will also show how recent developments in deep learning and large-scale datasets have propelled progress in this field. My goal is to illuminate the complexities of applying speech recognition to singing, evaluate current capabilities, and outline future research directions.
\end{abstract}

\section{Introduction}

Automatic Speech Recognition (ASR) technology has seen significant advancements in various domains but remains relatively underexplored in the field of singing. This gap in research may initially appear justifiable, given the widespread availability of song lyrics. However, the importance of ASR in singing extends far beyond mere convenience. It is crucial in niche music genres and less mainstream tracks, often described as 'long tail' music, where lyrics are not readily available. This technology also aids in making music more accessible to the hearing impaired through real-time captioning and can be an engaging tool for language learning and cultural exchange.

The necessity for this technology becomes more pronounced in the context of world music, a genre characterized by its diverse linguistic and cultural roots. Here, speech recognition can play a crucial role in breaking down language barriers and promoting cultural understanding. It can also enhance karaoke and entertainment applications, improve music search and discovery, and aid in music analysis and research.

Moreover, the future of AI models in this domain promises a more efficient approach to lyrics transcription. Current methods, often relying on user submissions or manual transcription, are time-consuming and can be inaccurate. Advanced speech recognition could potentially automate this process, providing quicker and more accurate transcriptions, and assist in content moderation for public performances.

The scope of ASR in singing extends beyond transcription. It includes practical applications like synchronizing lyrics with audio, identifying songs from sung lyrics (useful in platforms like music identification apps), transcribing lyrics from short audio clips, identifying the language of the song, and enhancing music recommendation systems by understanding lyrical content and sentiments.

My work in ASR for singing began in 2011, a time when the field was just starting to embrace new solutions for its challenges. This period also marked the rise of deep learning methods in various scientific areas, significantly impacting the field of ASR. My research coincided with this shift from traditional feature-based methods to more advanced, data-driven neural networks. The transition was particularly notable in ASR for singing. Earlier models, which were primarily designed and trained for regular speech, often underperformed with singing due to a lack of specialized adaptation techniques and data. The newer deep learning models have proven to be more effective in this regard. We are now beginning to see solutions that are not only theoretically sound but also practically viable for addressing the unique challenges of ASR in singing.

This paper aims to provide an overview of speech recognition specifically tailored to singing, including my own forays into the topic. It will begin with an examination of the unique challenges that singing poses as a data source, such as varied vocal styles, pitch variations, and background music interference. Following this, the paper will delve into specific research areas: phoneme recognition in singing (critical for understanding lyrics), sung language identification (which can be particularly challenging given the musical context), keyword spotting in songs (useful for searching and categorizing music), and methods for retrieving songs based on the sung lyrics. Additionally, we will review the progress in the complete transcription of songs, assessing current capabilities and limitations. In most cases, I will give a historical overview of first approaches, my own work between 2011 and 2018, and recent developments.

Finally, the paper will conclude with a summary of the key findings and a discussion on future research directions. This will include potential technological advancements, the integration of these systems into consumer applications, and the exploration of new use cases in the rapidly evolving landscape of music and technology.

\section{Singing as a speech data source}

Singing poses several unique challenges for speech recognition compared to normal speech. These challenges necessitate adapting existing speech recognition algorithms \cite{humphrey2018signal}:

\begin{description}[\compact]
\item[Larger Pitch Fluctuations] Singing involves more significant pitch variations than speaking, along with different spectral properties.

\item[Increased Loudness Variability] Loudness in singing fluctuates more than in speech.

\item[Pronunciation Variation] The musical context can lead singers to pronounce sounds and words differently compared to normal speech.

\item[Time Variations] In singing, sounds may be elongated or shortened to fit the musical rhythm, leading to more significant variations, especially with vowels. This was confirmed by a study comparing standard deviations of phonemes in speech and singing datasets.

\item[Different Vocabulary] Lyrics in songs often use different words and phrases compared to everyday conversation, with a focus on emotional topics.

\item[Background Music Interference] In polyphonic recordings, the presence of harmonic and percussive instruments adds spectral components that confuse speech recognition algorithms. Source separation algorithms ould be utilized to remove these components, but they are not always effective and can introduce artifacts. Vocal Activity Detection can be used to at least discard non-singing segments in songs, but it often makes errors in problematic cases for speech recognition, such as instrumental solos. Due to these challenges, most works focus on unaccompanied singing and leave the pre-processing as a separate step to be researched. Alternatively, algorithms are being developed with robustness to these influences in mind from the start.
\end{description}

Historically, ASR for singing faced significant challenges due to a lack of available data. Unlike speech data, which is often created specifically for ASR development, music and lyrics are generally subject to copyright restrictions that limit their use. Additionally, phoneme- or word-level annotations, crucial for ASR training, are typically scarce, and funding for creating such annotations is limited. As a result, early ASR models for singing were primarily adapted from speech data models and tested on small, available singing datasets.

When I began my research in this field, only two modest-sized datasets were available, each containing around 20 English-language pop songs with line-wise lyric annotations \cite{hansen2012recognition,mauch2012integrating}. I then utilized a dataset from \textit{Smule's} amateur karaoke app \textit{Sing!}, known as \textit{DAMP}, which comprised unaccompanied singing. Initially lacking lyric labels, I compiled lyrics from the \textit{Smule} website and conducted forced alignment using speech-trained models, which led to the creation of a set with 300 songs, each having 20 recordings \cite{kruspe2016bootstrapping}. Subsequent work improved the phonetic annotations of this dataset and expanded its scope \cite{dabike2019automatic}, making it a standard resource for ASR in singing.

The availability of data has significantly improved since then. New datasets have been introduced, such as \textit{MUSDB} \cite{schulze2021phoneme}, containing 150 mostly English-language songs with line-wise lyrics annotations, and \textit{vocadito} \cite{bittner2021vocadito}, featuring 40 manually labeled multilingual recordings. Another notable dataset is \textit{DALI} \cite{meseguerbrocal2018dali, meseguerbrocal2020creating}, which includes thousands of polyphonic recordings in various languages, with lyrics annotations obtained through semi-automated methods.

The recent proliferation of these datasets makes direct comparison with older approaches challenging due to the lack of established benchmarks. Therefore, in the following sections, my focus will be on exploring individual ideas for the tasks discussed, rather than providing numerical comparisons.

\section{Phoneme recognition for singing}\label{sec:phoneme}
Phoneme recognition in singing, also known as acoustic modeling, has traditionally been more challenging than in speech. This is due to the unique characteristics of singing described above, such as varied pitch and rhythm. For a long time, phoneme recognition was foundational for other ASR tasks like alignment, making it a critical area of study. Recognizing phonemes in singing is also a complex task for humans, as noted in \cite{hollien2000perceptual}.

Early phoneme recognition systems, which often relied on Mel-Frequency Cepstral Coefficients (MFCCs) and assumed pitch invariance, faced difficulties in accurately processing singing. Two initial systems using Hidden Markov Models (HMMs) were introduced in \cite{wang2003automatic, hosoya2005lyrics}. These were followed by approaches involving adapted Gaussian Mixture Model-HMMs \cite{mesaros2010automatic} and systems utilizing chorus repetitions \cite{mcvicar2014leveraging}. However, a lack of benchmark datasets for singing made direct comparisons with these early systems challenging.

The scarcity of large-scale, singing-specific training data also hindered progress. My own research initially involved using speech datasets like \textit{TIMIT} \cite{garofolo1993darpa}, but resulted in high error rates. To improve model robustness, I experimented with augmenting speech data to mimic singing characteristics, such as pitch shifting and time stretching \cite{kruspe2015training}. This led to reduced error rates and later informed the integration of these techniques into advanced models using Transformer architectures \cite{zhang2021pdaugment}.

Subsequently, I worked with the \textit{DAMP} dataset, derived from \textit{Smule Sing!} karaoke app recordings. By creating phoneme labels through forced alignment with TIMIT models, I demonstrated that direct training on singing data was significantly more effective \cite{kruspe2016bootstrapping}.

Presently, the focus in singing ASR has shifted towards full transcription, which involves integrating acoustic models with language models or employing end-to-end systems (also see section \ref{sec:transcription}). Nonetheless, the insights and methodologies from phoneme recognition research continue to be valuable for understanding singing's unique phonetic characteristics.

\section{Sung language identification}
Several methodologies have been explored for language identification in singing. In 2004, an unsupervised clustering approach was developed to create language-specific codebooks from input features, achieving an accuracy of 0.8 for English and Mandarin songs \cite{tsai2004towards}. \cite{schwenninger2006language} in 2006 utilized MFCC features for direct language model training, but faced challenges in singing and polyphonic contexts, indicating the complexity of the task . \cite{mehrabani2011language} in 2011 combined phoneme recognition with prosodic tokenization, testing on a multilingual corpus and achieving accuracies up to 0.83. In the same year, \cite{chandraskehar2011automatic} analyzed audio and video features in music videos, noting that identifying European languages was more challenging than Asian and Arabic languages, with accuracies around 0.45 using audio and 0.48 with both audio and video.

My work began with systems based on traditional audio features like MFCC and RASTA-PLP, fed into machine learning models \cite{kruspe2014gmm}. I later incorporated the i-vector technique, primarily used in speaker recognition, for feature reduction \cite{kruspe2014improving}. A notable observation was the models' tendency to confuse speaker characteristics with language features. Subsequently, I employed phoneme statistics for language identification, building on the phoneme recognition approaches described above \cite{kruspe2016phonotactic}. The phonotactic approach was subsequently also taken in \cite{renault2021singing}.

Recent developments have seen the adoption of advanced neural networks, along with the integration of auxiliary textual data such as song, artist, and album names \cite{choi2021listen}. The availability of large multilingual datasets has also improved, easing the research process \cite{santana2020music4all}. Despite these advancements, language identification in singing remains a challenging task, particularly for languages with limited resources in singing data.

\section{Keyword spotting}
Keyword-based search systems play a crucial role in various music-related applications, such as song discovery based on topics, playlist creation, similarity searches, genre classification, and mood detection. Early methods often relied on supplementary information like textual lyrics. For instance, a 2008 study employed vocal re-synthesis with MFCCs and power features for phoneme recognition, using Viterbi decoding alongside keyword-filler HMMs \cite{fujihara2008hyperlinking}. In 2016, the "LyricListPlayer" system utilized lyrics-to-audio alignment for keyword detection, incorporating NLP techniques for topic modeling \cite{nakano2016lyriclistplayer}. \cite{dittmar2012towards} presented an approach involving Statistical Sub-Sequence DTW for keyword spotting, which required audio recordings of key phrases. Additionally, \cite{dzhambazov2015searching} developed a score-aided method that combined acoustic keyword spotting with Sub-Sequence DTW and Dynamic Bayesian Network HMMs, tested on Turkish Makam music.

In my research, I focused on detecting arbitrary keywords in singing without needing extra information. The primary approach involved using keyword-filler HMMs, which consist of several phoneme-level states for detecting the desired keyword and an additional state for all other sounds \cite{kruspe2014keyword}. This method was further refined by incorporating knowledge about plausible phoneme durations to eliminate unlikely candidates \cite{kruspe2015keyword}.

Recently, keyword spotting in singing has seen less research focus, possibly due to the anticipation that comprehensive transcription systems may render individual keyword searches obsolete (refer to section \ref{sec:transcription}). Nevertheless, in scenarios where complete lyrics are unavailable or transcription systems are not fully accurate, keyword spotting methods retain their relevance. They offer the ability to identify more potential keyword instances within songs under model uncertainty.

\section{Lyrics alignment}\label{sec:alignment}
Lyric-to-audio alignment has been a more extensively researched topic compared to other areas discussed earlier. \cite{loscos1999low} first adapted speech recognition methods for singing using MFCCs and a modified Viterbi algorithm in a HMM for unaccompanied singing in 1999. However, this early approach was limited by its small database and lack of quantitative results. In 2004, a significant advancement was made with "LyricAlly," providing line-level alignments in polyphonic recordings, employing a mix of rhythm structure analysis, chord analysis, chorus detection, Vocal Activity Detection (VAD) through HMMs, and lyric segmentation \cite{wang2004lyrically}.

Further progress was made in 2006, focusing on syllable-level alignment and adapting speech acoustic models for singing. This period saw methods like auto-regressive HMMs for modeling high-pitched signals and MFCC-based Viterbi alignment with accompaniment sound reduction and phoneme model adaptation \cite{chen2006popular}. In 2010, \cite{mauch2010lyrics} introduced the use of chord labels to improve alignment accuracy \cite{mauch2010lyrics}. Other specialized approaches included a 2007 method for Cantonese singing, utilizing prosodic information from lyrics \cite{wong2007automatic}, and  structural analysis of song recordings in 2008 \cite{lee2008segmentation}. \cite{mesaros2008automatic} in 2008 incorporated harmonic re-synthesis for vocal separation, and \cite{gong2015real} in 2015 focused on vowel alignment in score-following algorithms. \cite{dzhambazov2016use} in 2016 enhanced accuracy by integrating note onsets into the alignment algorithm.

In my work, I applied Dynamic Time Warping (DTW) to align phoneme posteriorgrams, derived from phoneme recognition, with binary templates generated from lyrics. This approach was further refined with insights into phoneme probabilities and confusions, enhancing alignment accuracy. In the 2017 \textit{MIREX} challenge, this method achieved the lowest mean error rates in both unaccompanied and polyphonic music \cite{kruspe2017lyrics}.

The annual \textit{MIREX} challenge has improved the reproducibility and comparability of alignment methods. Recent developments include enhancements to acoustic modeling \cite{gupta2019acoustic}, an end-to-end solution using Wave-U-Net and CTC loss \cite{stoller2019end}, extensions to multilingual data \cite{vaglio2020multilingual}, and a computationally efficient approach for detecting anchor points \cite{demirel2021low}. Recently, the focus has shifted towards combining alignment with full transcription (see section \ref{sec:transcription}).

\section{Lyrics-based retrieval}
Lyrics-based retrieval, a relatively underexplored area in ASR research, involves identifying the correct textual lyrics and corresponding songs from a sung query. This technology is particularly beneficial for karaoke systems and voice-based search applications.

In 2006, \cite{suzuki2006music} developed a phoneme recognition system for lyrics retrieval. 
Their experiments showed that using a five-word query improved the retrieval rate, and integrating melody recognition further increased its accuracy. The query-by-singing system from \cite{wang2010improved} combined melody and lyrics in 2010,
while \cite{mesaros2010recognition} developed a system focused solely on lyrics for word recognition in singing. 

In my research, I utilized the outcomes of the phoneme recognition process described above, bypassing additional language modeling or melody integration, on a database containing 300 songs \cite{kruspe2016retrieval}. Sung lines from these songs served as retrieval queries. Adaptations made to the alignment process, as detailed earlier (section \ref{sec:alignment}), were instrumental in accommodating variations in recognized phonemes and common confusions \cite{kruspe2018retrieval}.

Recent years have seen limited focus on lyrics-based retrieval research. Full transcription capabilities could potentially address this task by enabling a fuzzy text search within a lyrics database. Nevertheless, integrating these lyrics-based methods with audio-based song identification could significantly enhance song search capabilities, especially for cover versions or queries by amateur singers who recall only fragments of the melody and lyrics.

\section{Transcription}\label{sec:transcription}
Lyrics transcription has long been viewed as the "holy grail" in the field of ASR for singing. Traditionally, an ASR pipeline involved an acoustic model to determine phoneme likelihoods and a language model to deduce the most probable sequences of phonemes and words, informed by statistical information from text data sets. However, in the context of singing, as noted in section \ref{sec:phoneme}, the acoustic models initially struggled to provide sufficient accuracy for full transcription. \cite{tsai2018transcribing} introduced deep learning-based (TDNN-LSTM) acoustic models trained on limited singing data in 2018.

With the availability of larger datasets and advancements in ASR technology, the pursuit of lyrics transcription has intensified recently. \cite{gupta2019acoustic} explored an end-to-end model combining acoustic and language modeling, although it initially underperformed compared to separate models. \cite{demirel2020automatic} then employed Time Delay Neural Networks (TDNNs) integrated with language models trained on diverse lyrics data, achieving improved error rates. Their subsequent work adopted a multistream approach for further enhancements \cite{demirel2021mstre}.

More recently, end-to-end transcription has become feasible with the adoption of Transformer and Conformer architectures. For instance, \cite{gao2022genre} extended these models with genre-specific adapters. \cite{ou2022transfer} successfully adapted wav2vec embeddings for singing, marking a significant leap in model performance. In the latest development, a novel approach involved performing ASR on music audio with the \textit{Whisper} system, followed by post-processing using \textit{ChatGPT}, leading to further reductions in error rates \cite{zhuo2023lyricwhiz}. It is anticipated that future research will increasingly leverage such versatile, pre-existing systems or adapt large language models (LLMs) to this domain.

\section{Conclusion and future work}

This paper has examined the challenges and recent progress in speech recognition for singing. We have seen how singing's unique characteristics, like varying pitches and durations, complex pronunciations, and background music, present different challenges from regular speech recognition. Significant advances have been made in key areas such as phoneme recognition, sung language identification, keyword spotting, and full song transcription. In recent years, improvements are largely due to advancements in deep learning and the availability of diverse, large datasets.

The transition from feature-based methods to deep learning signifies a major change in this field. These advanced models are better at understanding the subtleties of singing, leading to more precise and reliable ASR systems. Research exploring various languages and music styles continues to expand, enriching our understanding of music from around the world.

The potential impact of these developments on music discovery and recommendation systems is substantial. More accurate song identification and transcription can lead to better, more varied music suggestions, helping users discover new artists and genres. This not only improves the listening experience but also supports lesser-known music. Additionally, these advances in ASR can make music from different cultures more accessible and help overcome language barriers.

Looking forward, the focus is likely to shift more towards complete transcription of songs. The success in transcription could indirectly solve related tasks such as keyword spotting and lyrics alignment, offering a unified solution to several challenges in the domain. Moreover, the advent of large language models (LLMs) and Foundation Models, especially multimodal ones, presents a new frontier in ASR research. These models hold the promise of revolutionizing the field by providing more generalized and adaptable solutions.


\newpage
\bibliographystyle{essv}
\setlength{\bibsep}{8.0pt}
\bibliography{essv}

\end{document}